\documentclass[%
 reprint,
 prb,
 amsmath,amssymb,
 aps
]{revtex4-2}
\usepackage{hyperref}
\usepackage{hyperref}
\usepackage{graphicx}
\usepackage{bm}
\usepackage{enumitem}
\usepackage{amsmath}
\usepackage{amsfonts}
\usepackage{amssymb}
\usepackage{amsthm}
\usepackage{xcolor}
\usepackage{multirow}
\usepackage{booktabs}
\begin{document}
\preprint{APS/123-QED}
\title{Spin-orbit torque effects in all-electrical spin-wave spectroscopy experiments}
\author{J. Solano C\'ordova$^{1}$}
\email{jose.solano@fzu.cz}
\author{J. Mencos$^{3,4}$}
\author{R. Schlitz$^{2}$}
\author{Q. Rossi$^{4}$}
\author{D. Reustlen$^{2}$}
\author{J. Ali$^{1}$}
\author{S. Sailler$^{2}$}
\author{M. Lammel$^{2}$}
\author{M. Bailleul$^{4}$}
\author{D. Stoeffler$^{4}$}
\author{F. Casanova$^{3,5}$}
\author{S. T. B. Goennenwein$^{2}$}
\author{H. Reichlova$^{1}$}
\email{reichlh@fzu.cz}
\affiliation{$^{1}$ Institute of Physics, Czech Academy of Sciences, Cukrovarnická 10, 16200 Praha 6, Czech Republic}
\affiliation{$^{2}$ Department of Physics, University of Konstanz, 78457 Konstanz, Germany}
\affiliation{$^{3}$ CIC nanoGUNE BRTA, 20018 Donostia-San Sebastián, Basque Country, Spain}
\affiliation{$^{4}$ Departamento de Polímeros y Materiales Avanzados: Física, Química y Tecnología, University of the Basque Country (EHU), Donostia-San Sebastián 20018, Basque Country, Spain }
\affiliation{$^{5}$ IKERBASQUE, Basque Foundation for Science, 48013 Bilbao, Basque Country, Spain.}
\affiliation{$^{6}$ Institut de Physique et Chimie des Matériaux de Strasbourg, UMR 7504 CNRS, Université de Strasbourg, 23 rue du Loess, BP 43, 67034 Strasbourg Cedex 2, France}

\begin{abstract}
Electrical excitation and detection of propagating spin waves are key elements of magnonic devices. While conventional antennas consisting of metals such as Au or Cu excite spin waves predominantly through radio-frequency (RF) Oersted fields, antennas comprising Pt can additionally generate spin-orbit torques (SOTs) via the spin Hall effect. Recent theoretical work has proposed RF-SOT as an alternative mechanism for coherent spin-wave excitation, predicting distinct characteristics compared with conventional inductive excitation. Here, we investigate Pt antennas for the excitation of coherent spin waves in 105-nm-thick epitaxial $\text{Y}_3\text{Fe}_5\text{O}_{12}$ (YIG) films through two potential mechanisms: an oscillating Oersted field and an oscillating spin current. The raw experimental transmission spectra are strongly affected by fabrication-induced modifications to the YIG film associated with the etching process together with spatially inhomogeneous interface conditions induced by Pt. After removing these contributions, we find a remaining two wave propagation consistent with a combined excitation from Oersted fields and SOTs.

\end{abstract}
\maketitle
\section{\label{sec:level1_intro}Introduction\protect}
Spin waves are collective excitations of the magnetic order, whose quanta are known as magnons. They are promising carriers of information without the drawback of Joule heating from charge transport~\cite{kruglyak_magnonics_2010, flebus_magnonics_2024}. Furthermore, coherent spin waves exhibit a well defined phase relation between the dynamic magnetization components transverse to the saturation magnetization, enabling control of their amplitude and phase through oscillating magnetic fields.


Although novel approaches for the excitation of coherent spin waves have been developed~\cite{dieterle_coherent_2019, wojewoda_observing_2023, miedaner_excitation_2024, gallardo_coherent_2024}, inductive excitation and detection using microwave transducers remain the standard approach in propagating spin-wave spectroscopy, offering an all-electrical technique directly compatible with potential on-chip magnonic applications. The transducer converts high frequency electrical current (GHz range) provided by an external source into an oscillating Oersted magnetic field. The latter exerts a torque on the magnetization inducing an oscillating component $\mathbf{m}(t)$. When the excitation frequency $\omega$ matches the resonance frequency of one of the natural oscillating modes, the sample absorbs microwave power thereby generating coherent spin waves with the same frequency as the source ($\mathbf{m}\propto e^{-i\omega t}$). Patterning the transducer in an adequate geometry~\cite{vlaminck_spin-wave_2010} gives the Oersted field also a spatial periodicity with wave vector $\mathbf{k}$ that in turn will be imprinted in the phase of the exited coherent spin waves ($\mathbf{m}\propto e^{-i(\omega t + \mathbf{k} \cdot \mathbf{r})} $) allowing them to propagate. In this way, the inductive excitation approach enables electrical control of the spin-wave frequency and wave vector.

The transducers are commonly fabricated from good conductors like Au, Al or Cu~\cite{vlaminck_spin-wave_2010, bracher_phase--intensity_2016, solano_cordova_spin_2026} whose low resistivity facilitates impedance matching to the standard 50$~\Omega$ impedance of radio frequency electronics. This is an important requirement in order to minimize power reflections, unintended losses, and guarantee the most efficient inductive power transfer to the magnetization dynamics. However, despite their widespread use in ferromagnetic magnonics, conventional Au antennas couple inefficiently to compensated magnetic order. This limits their application to materials with vanishing magnetization, in particular ferrimagnetic, antiferromagnetic or altermagnetic insulators, which are particularly attractive for high-frequency magnonics and can also host low-frequency magnon modes in the GHz range with very small magnetic damping~\cite{Afanasiev.2014,Fischer.2020, Wang.2023}.
Therefore, schemes for electrical transducers enabling on-chip magnonics beyond the inductive coupling alone are very atractive.

Spin Hall and/or spin orbit torques, well established in the field of spintronics, hold the potential for an alternative electric coupling scheme. The corresponding transducers are made from heavy metals (HMs) like Pt, Ta and W, which can transform an electric current into a spin current through the spin-Hall effect~\cite{Althammer.2013}. The coupling to the magnet occurs via spin current injection from the heavy metal layer. Importantly, this type of coupling is effective in different types of magnetically ordered materials~\cite{Althammer.2013, Fischer.2020, Qiu.2025, Leiviska.2025}: the spin current exerts a spin orbit torque in the spin system of the magnet even if the magnetization is vanishingly small~\cite{Fischer.2020,Leiviska.2025}. Many experiments are based on a DC electric current that leads to a spin accumulation near the HM/magnet interface which then dissipates into the magnet via torques leading to the creation of thermal magnons. Building upon this approach, RF spin--orbit torque (RF-SOT) has recently been theoretically proposed as an alternative method for the generation of coherent spin waves ~\cite{morrison_coherent_2024}. An RF charge current applied to a heavy-metal transducer such as Pt generates an oscillating spin current via the spin Hall effect, which exerts a spin--orbit torque on the adjacent magnetic layer and directly drives coherent spin waves at the excitation frequency. Compared with conventional inductive excitation, RF-SOT was predicted to provide substantially higher excitation efficiency, distinct mode selectivity and a characteristic phase offset of approximately $3\pi/4$ between SOT- and inductively excited spin waves~\cite{morrison_coherent_2024}.

In this work we experimentally test these predictions. We use Au-based and Pt-based transducers to excite coherent spin waves in thin films (105 nm) of $\text{Y}_3\text{Fe}_5\text{O}_{12}$ films. Our spin wave spectroscopy experiments reveal a non-trivial spin wave propagation in the Pt-based devices, which we asses via a frequency- and time domain analysis. We rely on micromagnetic simulations to understand how modifications of the magnetic anisotropy can lead to the characteristic signatures of the measured signal. Comparing data between our Pt- and standard Au-based antennas we identify a potential RF-SOT enhancement to the propagation signals.

\section{\label{sec:Samples and devices}
Materials and devices}
We use commercially available, 105nm thick, [111]-oriented  $\text{Y}_3\text{Fe}_5\text{O}_{12}$ (YIG) films~\cite{Matesy} grown by liquid phase epitaxy on (111) $\text{Gd}_3\text{Ga}_5\text{O}_{12}$ substrates of 0.5 mm thickness. Pt layers with a thickness of 60~nm were deposited by sputtering onto the YIG surfaces~\cite{Supplemental}. The Pt-devices were patterned by negative-resist electron-beam lithography, followed by Ar-ion etching to define the device geometry. For the reference Au-devices, positive-resist electron-beam lithography was used to define the structures, followed by metal [Ti(5~nm)/Au(60~nm)] evaporation and lift-off~\cite{Supplemental}.

\begin{figure}[ht]
\includegraphics[width=0.48\textwidth]{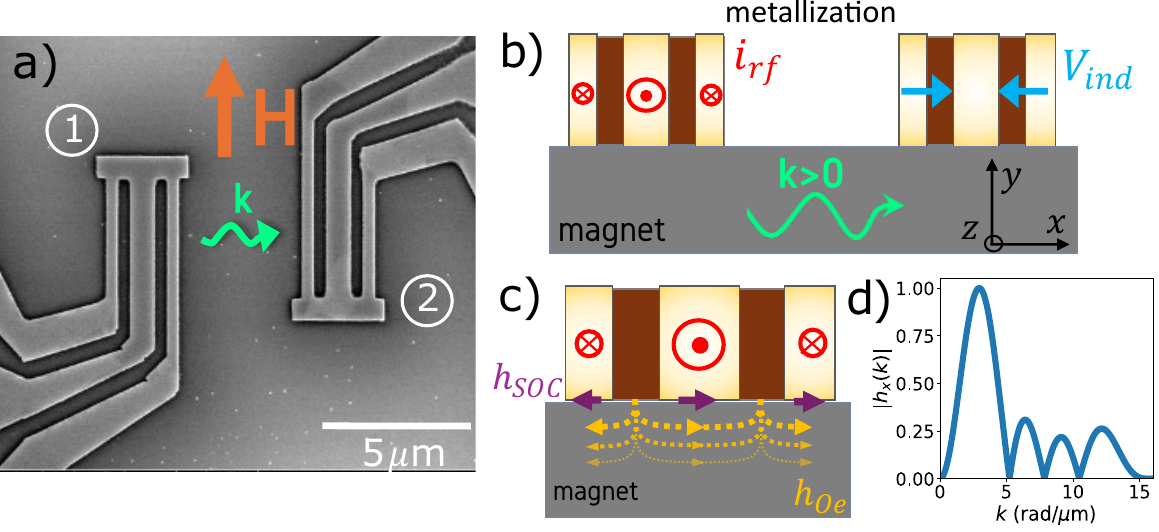}%
\caption{\label{fig:Device} (a) Scanning electron microscope image of an exemplary Pt-device. (b) Sketch of a device cross section displaying the excitation, propagation and transmission measurement. (c) Sketch of the two excitation magnetic fields $\mathbf{h}_\text{SOC}$ and $\mathbf{h}_\text{Oe}$ configuration at a given time. (d) Normalized Fourier transform (in reciprocal space) of the component along $x$ of the excitation field $\mathbf{h}=\mathbf{h}_\text{SOC}+\mathbf{h}_\text{Oe}$ near the interface of the metalization and YIG film. }
\end{figure}

The devices consist of a pair of coplanar waveguides, each is terminated in a constricted region, respectively, that act as lumped resistive inductor between the signal line and the ground pads (Fig.~\ref{fig:Device}). The constricted part is designed to shape the geometric period that defines the wavelength (wavenumber $k$) of the exited spin waves [Fig.~\ref{fig:Device} (c)]. For the same reason, these inductors also are selective filters for incoming inductive signals, i.e. these transducers act as antennas for the excitation and measurement of spin waves.  The nominal device have an edge-to-edge distance of $D=4\:\mu\text{m}$.

In reciprocal space, the antenna design yields a nominal main peak at $k_0=2.9\;\text{rad}/\mu m$ [Fig.~\ref{fig:Device} (d)]. The finite real space extension of the antenna results in a finite linewidth $\Delta k= 2.11$ rad/$\mu m$, as well as further less intense excitation peaks at higher wave vectors [Fig.~\ref{fig:Device} (d)]. The actual devices deviate only slightly from this nominal geometry due to the precision of the nanofabrication. This does not introduce important effects in the propagation, more details are discussed in the supplemental material \cite{Supplemental}.

While the geometry of the coplanar waveguide parts are designed for 50 $\Omega$ impedance, the antennas are more difficult to match due to their particular geometry. In the case of the Au-devices, a nominal $Z^\text{Au}=$50 $\Omega$ can be achieved by using the thickness combination Ti(5nm)/Au(51nm) with a resistivity $\rho_{Au}= 6.45\; \mu\Omega cm$. For the Pt-antennas, we only achieve an impedance of $Z^\text{Pt}=$129$\;\Omega$ for a 60 nm thickness and resistivity $\rho_{Pt}= 21.5\; \mu\Omega cm$. 

In order to assess the interface quality (spin transparency), we measured the spin Hall magnetoresistance ratios of our YIG/Pt samples and obtained $\Delta \rho/\rho =$6.5$\times 10^{-5}$ in agreement with previous characterizations~\cite{Velez.2016}. We furthermore used braodband ferromagnetic resonance to estimate a magnetic damping $\alpha=1.4\times10^{-4}$ for the bare YIG, $\alpha=3.8\times10^{-4}$ for the YIG/Pt films, and $\alpha=2.1\times10^{-4}$ for the YIG films after the etching of Pt (see further details in~\cite{Supplemental}).

\section{\label{sec:Experiment}Experiments}

\begin{figure}[ht]
\includegraphics[width=0.40\textwidth]{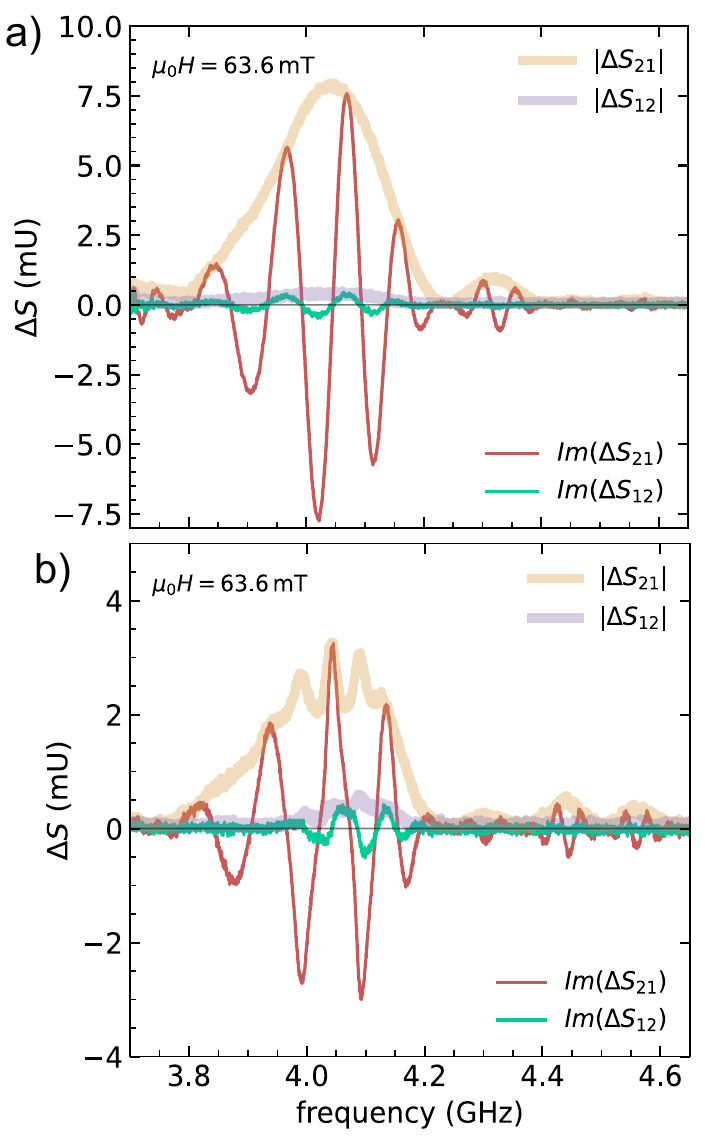}%
\caption{\label{fig:dL_0} Transmission signal between the two antennas with an edge-to-edge distance of 4 $\mu\text{m}$ at an applied field $\mu_0H=63.6 \text{mT}$ for (a) an Au-device (b) an Pt-device.}
\end{figure}

To quantify the spin wave propagation in a given device, we connect each of the two antennas to one port of a 2-port microwave vector network analyzer (VNA). The antennas are placed at the center of an electromagnet that provides an external magnetic field $\mathbf{H}$. This magnetic field is always applied in the plane of the devices and perpendicular to the direction of propagation of the spin waves [so-called Damon-Eschbach geometry~\cite{kostylev_non-reciprocity_2013}, Fig.~\ref{fig:Device} (a)]. The VNA is calibrated to measure the microwave scattering matrix $S$ between the two antennas.
$S_{21}$ ($S_{12}$) is the transmission measured at antenna 2 (1) due to excitation at antenna 1 (2) and in standard propagation it corresponds to a signal transmitted by spin waves with $k>0$ ($k<0$) [see Fig.~\ref{fig:Device} (a)]. At a given external magnetic field $H$, the VNA sweeps the frequency of the excitation current $i_{rf}$. When the frequency matches the resonance frequency $f_{k}$, the excitation microwave field of one antenna drives a strong precession of the spins (ferromagnetic resonance) resulting in coherent spin waves that propagate and can reach the second antenna. The propagating waves have a stray field that inductively couples to the latter, generating a voltage picked up as a change in the transmission between the antennas $S_{ij}$. To isolate only the transmission due to the spin wave resonance effect, we compute $\Delta S_{ij}(f)= S_{ij}(f,H)-S_{ij}(f,H_0)$, where $H_0>170 \text{mT}$ is an off-resonance reference magnetic field. The amplitude of $\Delta S_{ij}$ increases around the resonance frequency, while its complex phase oscillates as spin waves with slightly different wave vectors (different phase accumulation along the travel between antennas) are excited within the antenna's linewidth $\Delta k$ (see Fig.~\ref{fig:dL_0})~\cite{vlaminck_spin-wave_2010}. All our measurements are carrier out at room temperature.

Fig.~\ref{fig:dL_0} (a) shows typical experimental data, taken in a device with Au antennas (panel a), and with Pt antennas (panel b). For the Au-device, we observe a large difference between the amplitude of the waves propagating with $k>0$ ($\Delta S_{21}$) compared to that of the waves with $k<0$ ($\Delta S_{12}$). This is also known as amplitude non-reciprocity and is one of the hallmark properties of spin waves in the chosen Damon-Eschbach geometry~\cite{kostylev_non-reciprocity_2013}. Indeed, the measured ratio at resonance $|\Delta S_{12}|/|\Delta S_{21}|=0.05$ is in excellent agreement with that obtained by micromagnetic simulations (see Fig.~\ref{fig:Sims}). Interestingly the response $\Delta S$ of the Au-device [Fig.~\ref{fig:dL_0} (a)] is missing two of the higher resonances expected according to the nominal excitation field [Fig.~\ref{fig:Device} (d)]. This is an effect of the deviation from nominal geometry (see conductor broadening in~\cite{Supplemental}). 
Also, for the Au-device we observe that counter propagating waves have nominally the same resonance frequency (the maximum transmission is reached at around 4.04 GHz for both propagation directions). This is expected if the magnetic properties are homogeneous across the thickness of the film~\cite{Gladii.2016}. These properties and their agreement with our corresponding micromagnetic simulations show that our Au-devices act as ideal Oersted field-based transducers for the excitation/detection of propagating spin waves. Therefore, we will use them as reference for the discussion of the Pt-based antennas in the following.

The transmission spectra for the Pt-antennas [Fig.~\ref{fig:dL_0} (b)] show a distinctively different response. The first difference is an overall reduction of the amplitude of the transmission. Although at first sight this suggests the Pt-antennas transmit power less efficiently, we will see later that these devices are indeed comparable to the reference Au-devices in terms of coupling to the spin waves. Another difference is a modulation of the amplitude near the resonance frequency, which occurs for both $\Delta S_{21}$ and $\Delta S_{12}$. Despite this unusual feature, we observe an increase in the ratio of amplitudes at resonance $|\Delta S_{12}|/|\Delta S_{21}|=15-19\%$ (15$\%$ by comparing the minima of modulation, 19$\%$ comparing maxima). 
Last but not least, the secondary resonance peaks in the Pt devices resemble the nominal excitation spectrum [Fig.~\ref{fig:Device}(d)] more closely than those in the Au devices. We attribute this to the fact that the final Pt antennas fiducially reproduce the nominal geometry~\cite{Supplemental}.

\section{\label{sec:Analysis}Analysis}

To understand the spin waves excited and detected by the Pt antennas, we now examine the complex transmission signal in detail. We first demonstrate that several features of the measured response can be explained by interference effects due to modifications of the magnetic environment at the top surface of the film. After accounting for these initially unexpected contributions, we proceed to analyze the direct propagation signal.

\subsubsection{\label{sec:TimeD}
Time-gating}
\begin{figure*}[ht]
\includegraphics[width=1\textwidth]{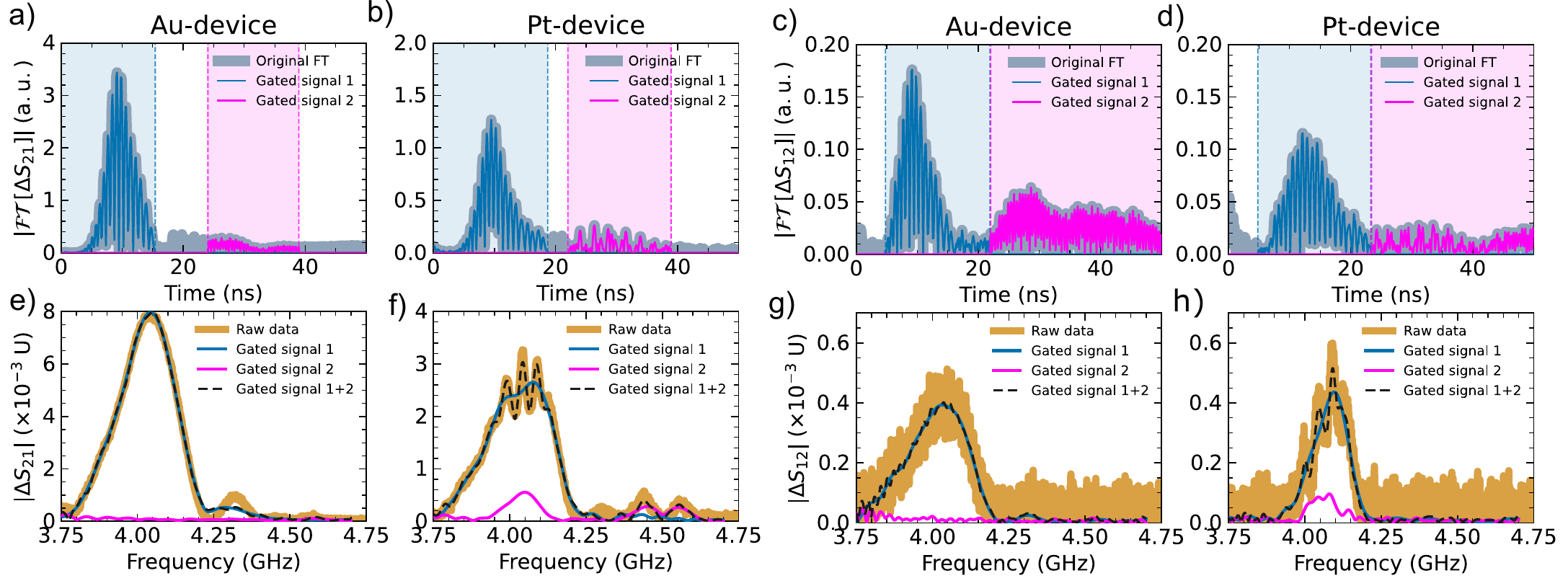}%
\caption{\label{fig:timeD} Fourier transform of the transmission signal $\Delta S_{ij}$ between the two antennas with an edge-to-edge distance of 4 $\mu\text{m}$ at an applied field $\mu_0H=63.6 \text{mT}$: for the Au-device (a) and (c), for the Pt-device (b) and (d). Inverse Fourier transform of the time domain gated signal plotted on top of the raw frequency spectrum (Fig.~\ref{fig:dL_0}) for the Au-device (e) and (f), for the Pt-device (g) and (h). Note the different vertical scales between the two types of devices for $\Delta S_{21}$.}
\end{figure*}

The shape of the main excitation peak of the Pt-devices suggests an interference between signals. To separate the contributions we resort to a Fourier transform into time domain. Following the approach described in reference~\cite{Devolder.2021} we calculate the Fourier transform $\mathcal{FT}[\Delta S(f)]$ of the scattering matrix for both type of devices and we plot them in Fig.~\ref{fig:timeD} (a-d).


For the Au-device the time domain spectra presents a single wave packet of large amplitude with maximum at 9.5 ns, and transmitting completely within the first ~16 ns for both propagating directions [Fig.~\ref{fig:timeD} (a) and (c)]. We time-gate \cite{Devolder.2021} the main peak inside a window (e.g. 0-15.5 ns) using a Heaviside function, and then calculating the inverse Fourier transform, allows to recover the frequency spectra shown in Fig.~\ref{fig:timeD} (e) and (g). In this way, we can isolate the main excitation peak of the Au-device in its totality for both propagation directions, showing a reciprocal propagation (amplitude non-reciprocity excluded). Indeed, if we consider the average time (9.5 ns) it takes for these waves to travel between the antennas and a calculated group velocity $v_\text{g}(k_0, \mu_0H=64mT)=0.63 \:\text{km/s}$~\cite{Supplemental}, then the resulting travel distance is $D_{eff}= D + 2D_0 = 6\: \mu m$, meaning that the spin waves travel an extra $D_0=1 \: \mu \text{m}$ below each antenna on top of the edge-to-edge distance $D=4\: \mu \text{m}$.

For the Pt-devices, the maximum in $\mathcal{FT}(\Delta S_{21})$ is also located at 9.5 ns (same as for the Au-device), but this peak has a tail up to longer times, meaning that part of the signal arrives with a certain delay. Additionally, the maximum for $\mathcal{FT}(\Delta S_{12})$ occurs at 12.1 ns, showing a strongly non-reciprocal propagation. To fully capture the main excitation peaks, we have to extend the gating windows up to 18.8 ns or 23.4 ns [Fig.~\ref{fig:timeD} (b) and (d)]. After calculating back the frequency spectrum from this gated signal [blue lines in Fig.~\ref{fig:timeD} (f) and (h)], we obtain spectra that resembles more closely that of the reference Au-device. Interestingly, $\Delta S_{21}$ has a two peak shape, while $\Delta S_{12}$ is a rather sharp single peak (even sharper than its corresponding signal in the Au-device). 

To recover the unusual modulation at the top of the Pt-device peaks, one has to include the part of the spectrum corresponding to longer times (3 times larger, i.e. gate 22-54 ns) than those of the main peak  [Fig.~\ref{fig:timeD} (b), (d), (f) and (h)]. In contrast, a similar gate in the case of the Au-device does not contribute with a significant signal around the main resonance frequency [Fig.~\ref{fig:timeD} (a), (c), (e) and (g)]. 

Based on these observations we attribute the unusual modulation of the transmission signal in the Pt-device to multiple reflections of spin waves between the antennas: A spin wave signal is generated in antenna 1, travels the distance $D_{eff}$ until antenna 2, part of it is measured at time $t_0$ (gated signals 1 in Fig.~\ref{fig:timeD}) and part of it is reflected, travels again to antenna 1 where it reflects once more and then back to antenna 2 where it is measured at a time roughly $3t_0$ (gated signal 2 in Fig.~\ref{fig:timeD}).

\subsubsection{\label{sec:freNonreciprocity}
Frequency domain}

By removing the interference contributions arising from multiple reflections, we isolated the part of our frequency spectra corresponding only to the direct propagation of spin waves between the antennas. We now proceed to analyize their properties. To fit the transmission signal we can use a Gaussian function~\cite{Haidar.2012}:
\begin{equation}
\begin{aligned}
    \label{eq:mutualInductance}
   \Delta S(f, \mathbf{k}) = \frac{A}{\delta} e^{ -\frac{(f-f_k)^2}{\delta^2} } e^{ i \left(\frac{2\pi }{f_{p}}(f-f_{k}) +\phi \right)},
\end{aligned}
\end{equation}
here $\mathbf{k}$ is the wave vector, $f_k$ is the central frequency of the Gaussian wave packet and the resonance frequency of the spin waves, $f_p$ corresponds to the oscillation period so that $\tau = 1/f_p$ becomes the time-of-flight for the propagating waves, $w$ is the width of the Gaussian wave packet, and $\phi$ is a generic phase. In Table~\ref{tab:gauss_fits} we summarize the results of the fits to the spectra from different devices.

In the case of the Au-device, there are no multiple reflections, so we analyze the raw spectra directly. By fitting Eq.~\eqref{eq:mutualInductance} to the main transmission peaks [see~\cite{Supplemental}], we get that these signals can be fully accounted for by a single spin wave mode with a small frequency non-reciprocity $f_{k>0}-f_{k<0}= 15\:\text{MHz}$, but essentially the same oscillation period $f_p= 0.105 \:\text{GHz}$, corresponding to an average time-of-flight $\tau = 1/f_p = 9.5\: \text{ns}$ which is in agreement with what we found in the time domain analysis above (Sec.~\ref{sec:TimeD}). As previously stated, the amplitude non-reciprocity at resonance amounts to $|\Delta S_{12}|/|\Delta S_{21}|=0.05$.

\begin{figure}[ht]
\includegraphics[width=0.42\textwidth]{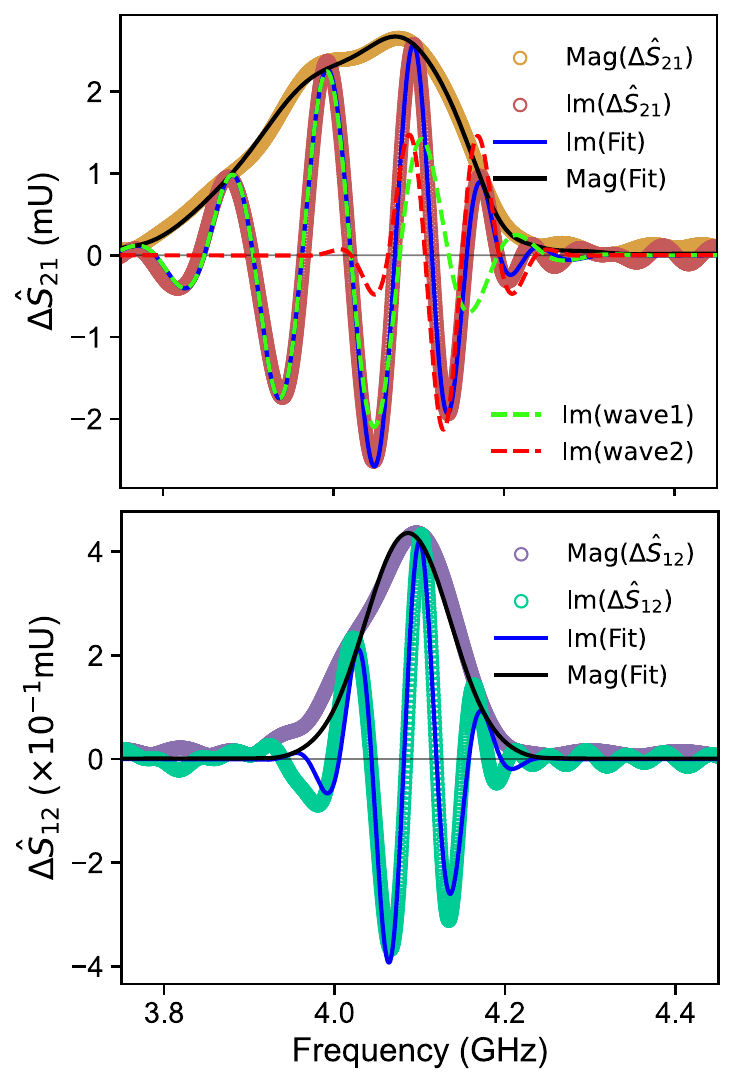}%
\caption{\label{fig:fits_dS} Inverse Fourier transform of the time-gated transmission signals of the Pt-device with $D=4 \:\mu \text{m}$ at $\mu_0H=63.6 \text{mT}$. The lines are the fits to Eq.~\eqref{eq:mutualInductance}: in the case of $\Delta \hat{S}_{21}$ we used a sum of two complex Gaussian terms (dashed lines), while for $\Delta \hat{S}_{12}$ a single one was used.}
\end{figure}

For the Pt-devices we analyze the inverse Fourier transform of the gated time spectra $\Delta \hat{S}$ in order to remove the effect of multiple reflections [gated signal 1 in Fig.~\ref{fig:timeD} (f) and (h)]. In Fig.~\ref{fig:fits_dS} we plot $\Delta \hat{S}$ and their respective fits. For $\Delta \hat{S}_{21}$ we need to use two different Gaussian packets [Eq.~\eqref{eq:mutualInductance}] to account for the double peak shape of the signal and the reduction of the oscillation period with increasing frequency. These waves have two different resonance frequencies separated by $121\:\text{MHz}$, they oscillate with different periods that translate into different times $\tau_{21}^1= 8.7 $ ns and $\tau_{21}^2= 11.5 $ ns. This second longer time explains the tail of the main peak we found in the time domain [Fig.~\ref{fig:timeD} (b)]. Additionally, by computing the phase difference between the two waves at resonance, we find $\phi_{21}^2-\phi_{21}^1=0.9(3\pi/4)$. In contrast, $\Delta \hat{S}_{12}$ can be fitted by a single Gaussian packet with a resonance frequency that sits between those of the two counter propagating waves. This wave has a shorter oscillating period equivalent to a time-of-flight $\tau_{12}= 13.15 $ ns. Finally, we observe that the amplitude non-reciprocity for this device amounts to an average of $|\Delta S_{12}|/|\Delta S_{21}|=0.2$.

Comparing the two type of devices, we find there is a large difference between $|\Delta S^{Pt}_{21}|/|\Delta S^{Au}_{21}|=0.3$ [Fig.~\ref{fig:timeD} (e), (f)]. This $70\:\%$ reduction is much larger than the expected $16\:\%$ decrease due to the larger impedance of the Pt-antennas ($Z^\text{Pt}/Z^\text{Au}=2.6$). Part of this extra reduction can be explained by the portion of transmission we removed with the time gating. In contrast, the counter-propagating signals in the Pt-devices have essentially the same amplitude as in the reference devices $|\Delta S^{Au}_{12}|/|\Delta S^{Pt}_{12}|=1$ [Fig.~\ref{fig:timeD} (g), (h)], despite a similar portion of signal removed by time-gating. 

Another point of comparison is the linewidth of the signals. Fig.~\ref{fig:timeD} (e-g) shows that the linewidths of the three main peaks are visually similar (see also Table~\ref{tab:gauss_fits}). However, $\Delta S_{12}$ for the Pt-devices [Fig.~\ref{fig:timeD} (h)] is exceptionally sharp. Both the linewidth and resonance frequency of this signal are closer to those of component $\Delta S_{21}^2$ (see Table~\ref{tab:gauss_fits} and wave 2 in Fig.~\ref{fig:fits_dS}).

Indeed, the similarities between $\Delta S_{12}$ and $\Delta S_{21}^2$ in the Pt-devices suggest that these two waves may result from a counter-propagating pair with similar origin (same spin wave mode, propagating with $k>0$ and $k<0$) that do not follow the usual amplitude non-reciprocity of the Damon-Eschbach modes~\cite{kostylev_non-reciprocity_2013}. In contrast, there is no signal for the counter-propagating wave of $\Delta S_{21}^1$. If this hypothetical wave would exist and follow the usual amplitude non-reciprocity, we would expect it to have an amplitude $\sim0.084 \: \times10^{-3}$, which is below our experimental noise level [Fig.~\ref{fig:timeD} (h)].

All these properties show that even after removing the multiple reflection contributions, the spin wave propagation in the Pt-devices do not follow the expected behavior of an standard inductive coupling between antennas and spin waves, hinting to an additional contribution to the propagation.

\begin{table}[ht]
\caption{\label{tab:gauss_fits}Gaussian fit parameters for each device and
propagation direction $\Delta S_{ij}$. $A$, $\delta$, $f_k$, $f_p$, are in units GHz and and $\phi$ in radians.}
\begin{ruledtabular}
\begin{tabular}{llccccr}
Device & Signal & $A/\delta\times10^3$ & $\delta$ & $f_k$ & $f_p$ & $\phi$ \\
\hline
\multirow{2}{*}{Au} & $\Delta S_{21}$   & 8.1 & 0.114 & 4.035 & 0.103 & 2.7  \\
                       & $\Delta S_{12}$   & 0.4 & 0.117 & 4.020 & 0.106 & 1.8  \\
                       \hline
\multirow{3}{*}{Pt} & $\Delta S_{21}^1$ & 2.3 & 0.141 & 4.008 & 0.115 & 2.5 \\
                       & $\Delta S_{21}^2$ & 2.1 & 0.068 & 4.129 & 0.087 & 4.7 \\
                       & $\Delta S_{12}$   & 0.4 & 0.071 & 4.086 & 0.076 & 0.4  \\
                       \hline
\multirow{2}{*}{Au-Etch} & $\Delta S_{21}$   & 14.1 & 0.121 & 4.040 & 0.106 & 1.6 \\
                       & $\Delta S_{12}$   & 1.3 & 0.086 & 4.002 & 0.100 & 0.0 \\

\end{tabular}
\end{ruledtabular}
\end{table}

\subsubsection{\label{sec:fabEffects}
Nanofabrication effect in the propagation}

We now discuss the origin of the multiple reflections. 

An important factor is the interfacial magnetic anisotropy induced by the Pt electrodes, which can modify spin-wave propagation in the YIG film. Previous studies have demonstrated that spin-orbit coupling in a nonmagnetic material can modify the magnetic environment of an adjacent magnetic layer, giving rise to an effective interfacial anisotropy~\cite{lee_2020,lee_2023}. Furthermore, transitions between regions of bare YIG and Pt/YIG bilayers have been shown to induce magnon reflections and even confinement~\cite{wang_current-controlled_2025}. 

To investigate whether such effects can explain our experimental observations, we perform micromagnetic simulations of spin-wave propagation in the presence of Pt-induced interfacial anisotropy. We model a 105 nm thick YIG film using standard parameters and realistic antenna excitation and detection area. We performed two types of simulations, the first where the YIG film is not modified (no anisotropy), and a second one (anisotropy) where we include an interfacial anisotropy term with constant $K_{\text{S}}=0.025 \: \text{mJ/m}^2$ \cite{lee_2023} only in the uppermost cells of the film's mesh that are in contact with the antennas' region (details in \cite{Supplemental}). 

\begin{figure}[ht]
\includegraphics[width=0.4\textwidth]{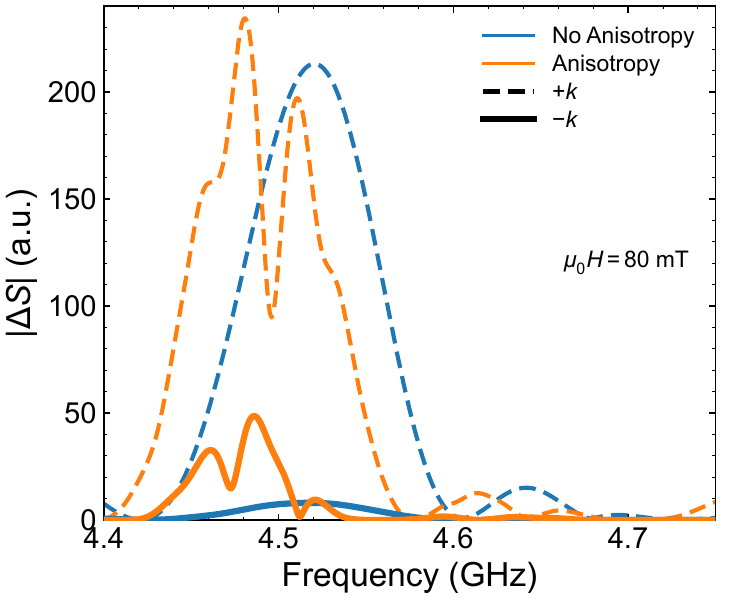}%
\caption{\label{fig:Sims} Simulated mutual inductance between two antennas with an edge-edge distance of 4 $\mu m$ at an applied magnetic field $\mu_0H=80 \text{mT}$. The simulation is performed for a 105 nm thick magnetic film with standard YIG parameters (no anisotropy) and one with interfacial magnetic anisotropy of constant $K_{\text{S}}=0.025 \: \text{mJ/m}^2$ localized only in the mesh cells of the film below the position of the antenna conductors.}
\end{figure}

We plot the results of the simulated transmission between the antennas in Fig.~\ref{fig:Sims}. Here we observe that without anisotropy we recover a transmission similar to that of the Au-devices, where there is a single peak and an amplitude non-reciprocity of $5\%$. After including the interfacial anisotropy, a large modulation of the transmission (multiple peaks) and an important increase of the signal for $k<0$ appears. We also observe a reduction of the resonance frequency compared to the case of no anisotropy.

Although the simulated spectra do not exactly reproduce the spectra we observe in the Pt-devices, they show that an interfacial anisotropy between the YIG and the Pt-antennas can produce qualitatively similar effects as those we have measured and identified with multiple reflections of the spin waves between the antennas.

Another possible source of modification to the surface of the films is the etching process we use to pattern the Pt-devices. To understand the effects of etching alone, we fabricated additional Au-etch-devices by the same lithographical process as the one used for the Pt-devices: we evaporated Ti(5nm)/Au(51nm) on a bare YIG film and then patterned the devices by Ar-ion etching. In this case, we do not expect the metalization to have an important role in modifying the YIG magnetic environment since it was shown that the effect of Ti is negligible~\cite{lee_2023} and our devices by lift-off do not show any modification. 

\begin{figure}[ht]
\includegraphics[width=0.48\textwidth]{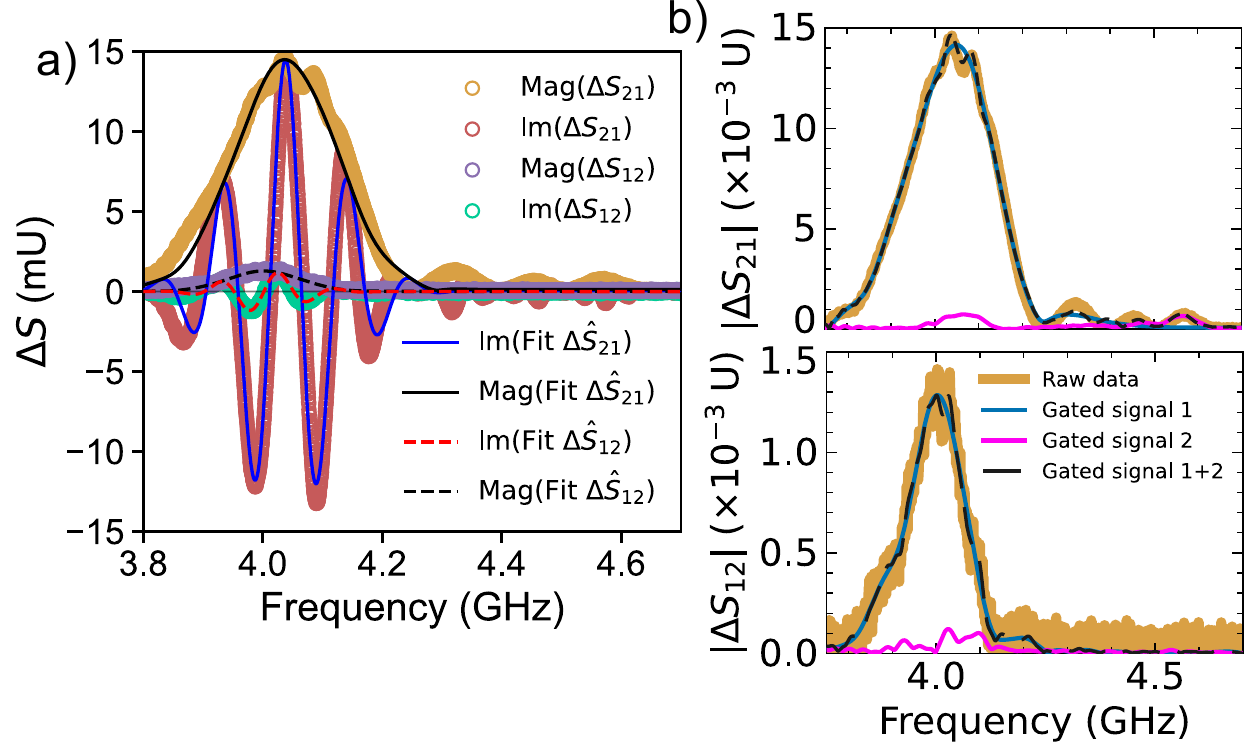}%
\caption{\label{fig:AuEtchedD} (a) Raw transmission signal $\Delta S$ between the two antennas with an edge-to-edge distance of 4 $\mu\text{m}$ at an applied field $\mu_0H=70 \text{mT}$ for an Au-etched-device. The lines are the fits to Eq.~\eqref{eq:mutualInductance} of the inverse Fourier transform of the gated transmission signal. (b) Inverse Fourier transform of the
time domain gated signals plotted on top of the raw frequency spectrum. Note the different scales of the two panels.}
\end{figure}

In Fig.~\ref{fig:AuEtchedD} we plot the transmission for this devices and the resulting inverse Fourier transform of the time-gated spectra. In the raw spectra we observe similar modulation features as the ones we found in the Pt-devices, albeit less pronounced, suggesting that the etching process alone can modify the YIG properties enough to produce the multiple reflections effect. Furthermore, by a similar process as in section~\ref{sec:TimeD}, we observe that by gating only the main transmission (gate 1) we recover a peak without modulations and that including longer times (gate 2) introduce them back [Fig.~\ref{fig:AuEtchedD} (b)]. Finally, we fit the gated spectra to Eq.~\eqref{eq:mutualInductance}, the resulting parameters are in Table~\ref{tab:gauss_fits} and we plot the results in Fig.~\ref{fig:AuEtchedD} (a). From the fits we can see that the waves in the Au-etch-device are similar to those in the reference Au-device, except for the amplitudes and a reduction for the linewidth of the $\Delta S_{12}$ signal. We hypothesize that both of these properties are linked to the better definition of antenna geometry achieved with etch-patterning which allows us to observe also the higher k-vector peaks of the excitation [Fig.~\ref{fig:Device} (d)]. In this case the ratio $|\Delta S_{12}|/|\Delta S_{21}|=0.09$ sits in-between those of the the other two type of devices.

This analysis demonstrates that the etching alone can also modify the spin wave propagation similarly to the way a inhomogeneous interfacial anisotropy can. Therefore, we expect both effects to contribute in our Pt-devices, explaining the pronounced multiple reflections.

\subsubsection{\label{sec:SOT}
RF-SOT contribution}
After disentangling the complex propagation signal in the Pt-devices by removing multiple reflections, fitting the remaining frequency components, and examining the role of Pt/YIG interfacial anisotropy or etching-induced modifications, one observation remains unexplained: the presence of two distinct propagating waves $\Delta \hat{S}_{21}^1$ and $\Delta \hat{S}_{21}^2$ we have identified in Fig.~\ref{fig:fits_dS} and Table~\ref{tab:gauss_fits}. The phase difference between these two contributions, which is very close to $3\pi/4$, suggests that they may originate from two different excitation mechanisms: conventional Oersted-coupling and the RF-SOT generation\cite{morrison_coherent_2024}. As discussed above, the components $\Delta \hat{S}_{21}^2$ and $\Delta S_{12}$ do not exhibit conventional amplitude non-reciprocity expected for Oersted field excitation. This further supports the RF-SOT origin of this part of the propagation, as this excitation mechanism is expected to excite spin waves reciprocally even in the Damon-Eschbach geometry~\cite{morrison_coherent_2024}. The fact that the overall transmission is still non-reciprocal is expected since our detection remains inductive only: the RF-SOT contribution excites spin waves with equal amplitudes in both directions~\cite{morrison_coherent_2024}, yet the detecting antenna still couples non-reciprocally to the stray field of these spin waves\cite{Supplemental}. Under this assumption, the effective RF-SOT contribution to the transmission would be a fraction $\sim A^2/A^1=0.45$ of the Oersted contribution.


\section{\label{sec:Conlusion}
Conclusions}

Motivated by theoretical predictions of RF-SOT as an alternative mechanism for coherent spin-wave excitation, we fabricated Pt- and Au-based antennas with identical nominal geometries on YIG thin films. The measured spin-wave transmission signals exhibit pronounced differences between the two types of devices. However, these differences are more complex than expected from the additional RF-SOT contribution alone. Through a combination of frequency- and time-domain analysis, supported by micromagnetic simulations, we identify multiple spin-wave reflections and modifications as important contributions to the measured response. We associate these effects to an inhomogeneous magnetic environment created by the magnetic anisotropy of the Pt/YIG interface and etching modifications to the surface of the films. Our results highlight the significant influence of modifications to the magnetic environment induced by interfaces and nanofabrication on propagating spin-wave spectroscopy using Pt antennas.

After accounting for these contributions, we identify a residual signal exhibiting a characteristic phase shift of $3\pi/4$ relative to the inductively excited spin waves, consistent with theoretical predictions for RF-SOT excitation. While further experiments would be valuable to conclusively establish the origin of the signal, we attribute these observations to coherent spin waves excited via RF-SOT.

\begin{acknowledgments}
We thank Sumit Gosh for theoretical discussion and Kim Gmeiner for assistance with nanofabrication. J.S. is co-funded by the European Union (Physics for Future – Grant Agreement No. 101081515). The work was also supported by the Dioscuri Program LV23025 funded by MPG and MEYS and TERAFIT - CZ.02.01.01/00/22$\_$008/0004594. F.C and J.M. acknowledge funding from MICIU$/$AEI$/$10.13039$/$501100011033 (Grant No. CEX2025$-$001634$-$M) and from MICIU$/$AEI and ERDF$/$EU (Project No. PID2024$-$155708OB$-$I00). We gratefully acknowledge funding
by the Deutsche Forschungsgemeinschaft (DFG, German Research Foundation) via projects ID 490730630 and ID 425217212 (SFB 1432), and technical support by the nano.lab facility of the University of Konstanz. J.M. acknowledges funding by the Department of Education of the Basque Government under the Predoctoral Programme for Training of Non-doctoral Research Staff (Fellowship PRE$\_$2022$\_$1$\_$0297).
\end{acknowledgments}
\bibliography{AuvsPtDevices}
\bibliographystyle{apsrev4-2}
\end{document}